# RESEARCH INTEGRITY AND GENAI: A SYSTEMATIC ANALYSIS OF ETHICAL CHALLENGES ACROSS RESEARCH PHASES

## A PREPRINT

Sonja Bjelobaba[1*], Lorna Waddington [2], Mike Perkins [3], Tomáš Foltýnek [4], Sabuj Bhattacharyya [5], Debora Weber-Wulff [6]

[1] Uppsala University, Sweden
[2] University of Leeds, UK
[3] British University Vietnam, Vietnam
[4] Masaryk University, Czechia
[5] Institute for Stem Cell Science & Regenerative Medicine, India
[6] HTW Berlin, Germany

[*] Corresponding Author: sonja.bjelobaba@uu.se

December 2024

# Abstract

**Background**

The rapid development and use of generative AI (GenAI) tools in academia presents complex and multifaceted ethical challenges for its users. Earlier research primarily focused on academic integrity concerns related to students' use of AI tools. However, limited information is available on the impact of GenAI on academic research. This study aims to examine the ethical concerns arising from the use of GenAI across different phases of research and explores potential strategies to encourage its ethical use for research purposes.

**Methods**

We selected one or more GenAI platforms applicable to various research phases (e.g. developing research questions, conducting literature reviews, processing data, and academic writing) and analysed them to identify potential ethical concerns relevant for that stage.

**Results**

The analysis revealed several ethical concerns, including a lack of transparency, bias, censorship, fabrication (e.g. hallucinations and false data generation), copyright violations, and privacy issues. These findings underscore the need for cautious and mindful use of GenAI.

**Conclusions**

The advancement and use of GenAI are continuously evolving, necessitating an ongoing in-depth evaluation. We propose a set of practical recommendations to support researchers in effectively integrating these tools while adhering to the fundamental principles of ethical research practices.

# Introduction

Artificial Intelligence (AI) has opened new possibilities in research, offering novel methods and enhancing data analysis. The meteoric rise of Generative AI (GenAI) has introduced transformative tools capable of creating realistic text, images, and even entire datasets. As the impact of GenAI on research could be significantly disruptive (European Commission Directorate General for Research and Innovation, 2024), the integration of these technologies in research practice warrants careful consideration of research ethics and integrity concerns. This consideration is particularly urgent, given that an estimated 1% of abstracts submitted to the preprint repository arXiv in 2023 contain evidence of GenAI tool usage (Gray, 2024), highlighting the rapid adoption of these tools in academic research. This is an area of concern, given the dramatic rise in the use of these tools within the scientific community (Kobak et al., 2024; Liang et al., 2024) and the ever-higher abilities of GenAI in research: there have even been demonstrations of GenAI tools which can carry out research with very limited human intervention (Sakana.ai, 2024).

The ethical implications of using GenAI in research are complex and multifaceted. While much of the discussion related to the ethical considerations of AI tools has focused on academic integrity concerns related to student use of AI tools (Cotton et al., 2024; Foltýnek et al., 2023; Perkins, 2023), there has been limited research exploring the use of GenAI in the context of academic research (Perkins & Roe, 2024b). The challenges of integrating GenAI into research include difficulties in replicating research findings due to variations in outputs across different iterations (Perkins & Roe, 2024c) and issues with transparency in how these tools are utilised (Weber-Wulff et al., 2023). Initial considerations have been published regarding the ethical issues behind the use of these tools in specific disciplines, including psychology (Chenneville et al., 2024), health research (Spector-Bagdady, 2023), and software engineering (Kirova et al., 2023), but there is no evidence of any empirical work identifying ethical challenges across a broad spectrum of research activities. However, several conceptual articles have explored the ethical challenges prevalent in the use of GenAI tools in research practice. For example, Resnik and Hosseini (2024) summarised the ethical challenges involved in using GenAI tools for research purposes, identifying the risks of bias, errors, the lack of moral agency present in AI tools, and the challenges of a lack of explainability of the output from these tools (the 'black box' problem). To overcome these challenges, Kurz and Weber-Wulff (2023) offer three fundamental rules for AI use: it must be permitted, made transparent, and authors must accept complete responsibility for the generated output.

Various multilateral organisations have attempted to address these challenges through guidelines and recommendations, with UNESCO having published reports warning of potential ethical risks while acknowledging potential benefits for research productivity (Miao & Holmes, 2023). Individual journals have also published guidance for scholars, although these policies have been criticised for their sometimes limited scope (Perkins & Roe, 2024a).

This study addresses two fundamental research questions concerning the ethical implications of GenAI in academic research. First, we examine what ethical concerns arise from the use of GenAI in different phases of research, from idea generation to dissemination of results. Second, we investigate how to support and encourage the ethical use of GenAI for research purposes. Through an investigation of a variety of GenAI tools, we examine the key ethical implications associated with GenAI use in research, including the protection of sensitive personal information, accuracy and reliability of outputs, bias and discrimination, transparency in tool usage, intellectual property rights, and scientific misconduct. We contribute to the literature by developing a set of practical recommendations that can support researchers in integrating these tools while maintaining the fundamental principles of good research practices.

# Methodology

The ethics of GenAI in research remain largely unexplored despite its rapid adoption across academic disciplines. Although GenAI tools proliferate and evolve continuously, comprehensive tool-by-tool evaluation is impractical. Therefore, this study adopts a strategic approach, examining representative tools at key research stages while using our previously published work as empirical benchmarks. Our investigation draws upon the *European Code of Conduct for Research Integrity* (Allea - All European Academies, 2023) as its guiding framework, leveraging our interdisciplinary team's expertise spanning history, computer science, ethics, and medicine to identify the systemic ethical challenges emerging from GenAI integration into research practices.

Our methodological framework encompasses the complete research lifecycle, beginning with conceptualisation and design. During this initial phase, researchers engage in ideation, hypothesis formation, and a comprehensive literature review. During the literature review process, researchers can utilise GenAI tools to facilitate cross-language research, enabling them to access and synthesise literature in unfamiliar languages. This phase also incorporates the development of funding proposals and the preparation of ethics reviews, each presenting unique considerations for GenAI integration.

The data processing and analysis phases present distinct methodological challenges and opportunities. Researchers have employed GenAI tools for data collection, processing, and sophisticated analysis. These tools prove particularly valuable for coding and categorisation processes, while also supporting programming and debugging in computational research. The transcription of audio data, such as research interviews, represents another crucial application. Additionally, the generation of images and videos for academic purposes introduces novel ethical considerations that require careful examination.

The development of content and communicating that content effectively forms another critical component of our methodology. Research writing rarely follows a linear trajectory, making GenAI tools valuable for text creation and refinement. These tools may particularly support researchers who are English as a Second Language (ESL) speakers facilitating more effective communication of scientific content. Visual content generation capabilities enhance research presentation, whereas academic writing tools improve clarity and structure. Each of these applications demands careful consideration of the ethical implications and potential impacts on research integrity. The dissemination and review phase concludes our methodological framework, encompassing peer-review processes, publication ethics, research communication, and impact assessment.

Our analytical approach employed detailed case reports derived from the practical applications of GenAI tools throughout these research phases. These reports systematically examined tool applications, compared outcomes with traditional research approaches, identified potential ethical challenges, and assessed implications for research integrity. While acknowledging the existence of numerous additional applications of GenAI in research, our methodology focuses on representative cases that illustrate key ethical challenges. This targeted approach enables a deeper exploration of fundamental issues rather than an exhaustive evaluation of specific tools. Throughout our analysis, we recognise that research processes often occur concurrently, rather than sequentially, introducing complex interactions between different phases and their associated ethical considerations.

This methodological framework supports the systematic investigation of GenAI's ethical implications while maintaining research integrity. By examining specific tools within their research context, comparing outcomes with established benchmarks, and analysing emerging ethical challenges, we provide a foundation for understanding and addressing the implications of GenAI integration in academic research. Our approach acknowledges both the opportunities and challenges presented by these tools while emphasising the importance of responsible integration that upholds academic integrity and research quality.

# Research Phases and AI Tools

## Formulating hypotheses, research questions and study design

Both general GenAI platforms like ChatGPT and specialised research tools such as Kahubi (Kahubi.com, 2023) have the ability to generate research questions and hypotheses for specific topics, with the ability of GenAI to handle large volumes of literature contributing to the possibility of identifying gaps in current literature, as well as creating novel hypotheses from the existing literature (Tong et al., 2024).

However, when testing the ability of these tools to support idea development, we identified significant limitations in their capabilities. For example, when prompted to identify research questions about country-specific climate change impacts, Kahubi categorised these potential impacts into ten themes (for example, Agricultural Vulnerability, Water Resources, etc.) and proposed some potential research questions. However, many of these questions are already well explored in the existing literature, suggesting limitations in identifying truly novel research directions. Moreover, Kahubi's literature summaries for existing literature closely mirrored the sentence structure of source abstracts, raising concerns about potential plagiarism if these summaries were used by future authors.

Additional ethical challenges include the risk of underlying structural biases influencing future research design, and the possibility that any GenAI tool's built-in guardrails might restrict academic freedom by limiting the exploration of certain topics in the proposed research questions. Although current research indicates that GenAI tools have a strong potential to produce research ideas which are more innovative than those produced by human researchers (Si et al., 2024), it is unclear whether these ideas can surpass the most innovative human ideas (Conroy, 2024). Overall, the key ethical problem remaining when considering how GenAI can support this area of the research process is the opacity of the GenAI tools (Cao & Yousefzadeh, 2023), leading to uncertainty about how any such ideas have been generated and what biases or limitations might be present in this process.

Despite these limitations, a survey of 1,600 researchers revealed that 32% of the respondents used GenAI tools to brainstorm research ideas (Van Noorden & Perkel, 2023), demonstrating the immediate impact of these tools on the future of scientific research.

## Literature Review

### Literature Gathering

At the start of any research endeavour, one needs to look for potential prior work on all or part of the question being addressed. Once the literature is found, either online or in a library, many researchers already have a pattern of bookmarking online sources and saving bibliographic information in a reference database. Many tools, such as Citavi, Mendely, Endnote, and Zotero, are commonly used and work well. It is expected that GenAI tools for literature gathering do not use a stand-alone database but integrate well with at least one, if not all, of the above-mentioned bibliographic databases.

We examined Perplexity, ResearchRabbit, Consensus, Elicit, and Litmap. A number of issues were identified in the systems evaluated. One of the most problematic issues was that much more online, non-academic literature was found than academic literature. Among those found were links to potential predatory journals that are, of course, not marked as such. The literature found was not always related to the given prompt. In one case, 90% of the literature "found" was not relevant to the prompt; they seemed to be the result of simple word searches and not the meaning of the multi-word terms. The systems also generated convincing, yet fabricated references, demonstrating issues with accuracy and reliability.

### Textual Understanding & Summarization

The tools in this category aim to summarise the information from one or more documents to either obtain an overview of the key findings in the given paper(s) or to create text that can be used directly in the manuscript, typically in the literature review section. This category overlaps with the literature-gathering stage. For example, Elicit, a tool designed to support literature review, not only provides a list of the most relevant papers but also a paragraph with a summary of the top four papers. It also provides an abstract summary in one sentence of each paper identified as relevant to the given query. Such an instant literature review is not based on critical thinking and may result in an increase in superfluous and non-relevant citations (Fong & Wilhite, 2017).

There are multiple tools for textual understanding and summarisation, such as Enago Read, SciSummary, Scholarcy, NotebookLM, and Resoomer. Moreover, universal AI tools such as ChatGPT can be used for this purpose with suitable prompts. In our testing, we explored the ability of ChatPDF (free version) and NotebookLM to provide information about a provided research paper (Foltýnek et al., 2020). While ChatPDF could provide a reasonable summary of key findings, it failed to answer simple questions about the authors of the study or the methodology. It also hallucinated the list of authors and claimed that the paper did not explicitly mention key results, despite these being included. NotebookLM was able to correctly answer the same questions. It also created a colloquial "deep-dive" podcast about the paper that was reasonably correct and provided a list of relevant concepts with explanations and some questions (with answers) about the paper.

Recent research has indicated that the accuracy of information extraction depends on the position of such information in the text. Liu et al. (2023) found that the performance of a large language model when processing long texts degrades significantly when relevant information is located in the middle of long contexts, with better performance for information at the beginning or end. However, this phenomenon has also been observed in humans, who tend to remember information from the beginning (primacy effect) and end (recency effect) of a list of items (Keith, 2013) or a longer document such as a novel (Copeland et al., 2009).

Another ethical concern revolves around intellectual property rights and copyright ownership. Many systems require users to confirm that they hold copyright for uploaded content, with some platforms even requesting users to upload their own bibliography and copies of papers relevant to the topic being explored. This practice raises serious copyright concerns, especially because the intended use of these materials by AI service providers often remains unclear in terms of service. In examining the terms and conditions of several tools, most of these required the transfer of some intellectual property rights to the service providers, insisting that the user confers the rights to use the material for further training of the systems. For the user to do so, they would have to hold this specific right, which is not normally granted to users, for example, when downloading papers from behind a library paywall. Given that very few users read the terms and conditions of the services they use (Bakos et al., 2014), and those who read them mostly merely skim through the text (Steinfeld, 2016), we conjecture that users, in general, may not be aware of such provisions and may inadvertently upload copyrighted content to these services. This issue is highlighted in Figure 1, taken from a journal homepage which clarifies that users do not hold the relevant rights to upload material taken from the journal to an external GenAI tool.

Figure 1: An example journal making the copyright issue clear

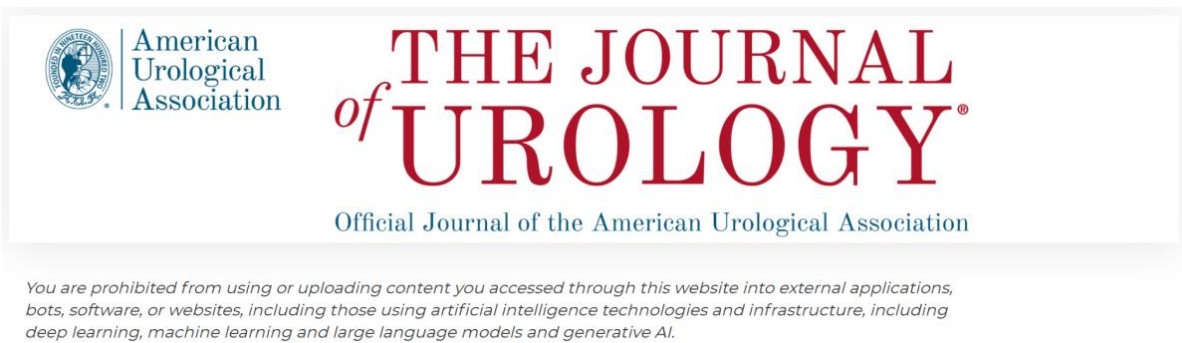

You are prohibited from using or uploading content you accessed through this website into external applications, bots, software, or websites, including those using artificial intelligence technologies and infrastructure, including deep learning, machine learning and large language models and generative AI.

To mitigate intellectual property infringement risks, users should use only services which do not require the transfer of intellectual property rights. To mitigate the risk of incorrect or misleading information, researchers should always verify the information obtained from AI-based information extractors and not rely on simple summaries of text generated by these tools. Therefore, we recommend the use of tools which link information provided by the tool to specific sections of text, thus allowing for easy verification and preserving the benefits of GenAI use, such as saving significant time reading or searching the document.

## Study Design and Data Collection

Prior to conducting any research, GenAI tools can theoretically be used to identify ethical problems in a study before conducting an ethical review. However, this approach itself introduces potential ethical issues that stem from the limitations of GenAI in comprehending human ethical values, understanding the nuances and cultural specificities of a particular context, potential power imbalances, and the inherited bias of the systems. The lack of transparency in how GenAI tools reach their conclusions can make it difficult to assess whether ethical challenges have been properly addressed. In addition, researchers should be aware that posting unpublished research instruments in a GenAI tool might lead to the inclusion of such research in a GenAI training set.

Many GenAI tools can be used to create surveys and interview questions. One of the main ethical problems in this area is that GenAI can perpetuate biases and lack sensitivity and contextual understanding, resulting in questions being discriminatory or skewed, assuming stereotypes, failing to accommodate different perspectives, or even inflicting harm and distress. This can lead to biased data collection. The ethical training of the system can also indicate whether the proposed questions are problematic; for example, explaining that questions related to the Bell Curve and IQ related to ethnicity might be sensitive before creating survey questions. Although such ethical training might be positive, it can interfere with the freedom of academic enquiry, censoring researchers regarding the types of questions that can be asked.

One of the main concerns in research ethics is the protection of human subjects (Council for International Organizations of Medical Sciences, 2016). Researchers have a duty to give research subjects relevant information about the research and the opportunity to provide consent to participate in that research. Such relevant information could include using GenAI in a way that can risk the autonomy of research participants (Perni et al., 2023). Currently, GenAI can be used to improve the text used to provide informed consent. However, owing to the risk of hallucinations, it is not suitable to use GenAI to generate informed consent statements without thorough human oversight (Currie et al., 2023; Shiraishi et al., 2024).

**Transcription**

In many disciplines, transcribing handwritten documents and audio interviews is tedious and error prone. Outsourcing transcription tasks has been a common practice, but little care has been given to human transcriptionists dealing with sensitive material (Wilkes et al., 2015). For example, in a project involving transcribing the testimonies of women who experienced the loss of two or more children during pregnancy, transcriptionists felt emotional distress or vicarious trauma, especially if the researchers failed to provide sufficient background information (Hennessy et al., 2022).

AI-based systems for transcription may be embedded in online meeting platforms, such as Zoom or MS Teams, or are available as stand-alone applications. They offer various functionalities, including real-time closed captioning and translation capabilities, and may overcome ethical issues related to human transcriptionists. However, there are other significant ethical implications of using such systems. The outputs may include hallucinations, or the transcription software may produce lower-quality results when transcribing the speech of certain groups. For example, the experiments of Blodgett and O'Connor (2017) showed that some systems perform more poorly when analysing the language of women and certain minorities, such as Afro-Americans. In case the interviews contain personal or sensitive information, using such systems may even violate data protection laws, for example, ensuring interviewees are aware of where their data is being stored and or processed. It must also be determined whether the audio recording or transcribed text will be used for future AI training, which could violate the privacy of the interviewees. These issues may be mitigated by running such transcription software locally, which does not involve any data transfer outside the research institute.

## Data Processing and Analysis

Although data processing can take different forms, in general, we are talking about activities aimed at preparing or transforming raw data into a form which is suitable for analysis. In many cases, raw research data is in a form which does not allow for analysis and thus requires transformation into a completely different form: transcription of audio recordings, data coding, or segmentation, for example. In other cases, data might be in a form that would allow for analysis, although further processing steps are needed, such as filtering, noise removal, imputation of missing values, normalisation, standardisation, resampling, feature extraction, and smoothing.

GenAI tools have the potential to support the processes of data analysis, although owing to their stochastic nature, particular care must be taken when evaluating the outputs (Perkins & Roe, 2024b). The use of GenAI tools for data analysis has a unique set of ethical considerations, which are distinct from other research phases. Although a small number of studies have demonstrated the use of GenAI tools in qualitative analysis (Bijker et al., 2024; Gao et al., 2023; Perkins & Roe, 2024a; Yan et al., 2024), no empirical studies have been identified as using GenAI tools for quantitative research (Perkins & Roe, 2024b). This section explores the potential ethical issues involved in using these tools to process and analyse data.

There are two distinct options for using GenAI for data analysis: First, the GenAI tool directly analyses the data—this applies mostly to qualitative analysis, but is also possible for quantitative analysis. Second, the GenAI tool provides advice about appropriate analytical methods, and researchers perform the analysis themselves using a suitable statistical tool. In particular, early career researchers may be unsure about what processing steps are suitable in a particular case, and following the advice of a GenAI tool may lead to improper methodology or even scientific misconduct. For example, imputing missing values may be methodologically correct in one context but may be considered data fabrication in another context.

### Data Anonymisation

Anonymisation of well-structured data is usually straightforward and includes the removal of personal identifiers and other attributes that may allow for linking particular data points to a specific person. In some cases, further steps, such as data aggregation, are necessary to ensure anonymisation.

Anonymisation of unstructured (textual) data often requires either a significant human workload or the use of NLP techniques or LLM-based tools. There are various tools that researchers can use to anonymise textual data such as Textwash (Kleinberg et al., 2022), but they are typically not adaptable to domains outside of those they were originally designed for, as the anonymity of a text document is largely influenced by its specific domain and context (Sotolář et al., 2021). The effectiveness of current anonymisation methods is further challenged by advances in GenAI, which can potentially de-anonymise individuals based on the remaining contextual information (Patsakis & Lykousas, 2023). This raises concerns regarding the balance between privacy protection and data usability. Nonetheless, these issues exist regardless of whether anonymisation is performed by humans or machines. Similar to other parts of the research process, if researchers use online tools, they should be aware of their terms and conditions. Transmitting personal or sensitive data to third parties without explicit consent is forbidden in most jurisdictions. Another risk is that the tool fails to locate all personal identifiers. However, the same risk also applies to human anonymizers, and a possible solution involves either a thorough check for forgotten personal data or a decision not to publish the dataset.

**Qualitative Analysis**

To explore the potential benefits of GenAI tools in supporting qualitative analysis, we explored an open source database of GenAI abstracts named the "AI in higher education database" (AIHE V1) (Ismail et al., 2024). This includes abstracts of 160 manuscripts related to GenAI and HE published between 20 November 2022 and 31 December 2023. In this test case, Claude 3 Opus was used to conduct an inductive thematic analysis on the database. This was done by uploading the raw file to Claude and asking for help in conducting a comparative qualitative thematic analysis. Claude was asked to explore the database and suggest research questions, inductively generate codes and subcodes, perform coding, identify themes, conduct sentiment and temporal analyses, and identify practical implications.

Although the GenAI tool demonstrated that it was able to accurately interpret user requests related to specific methodologies and rapidly create themes from the data, other issues emerged which were a cause for concern, specifically around the area of data interpretation and accuracy. This investigation revealed that GenAI tools can quickly generate codes, subcodes, and themes from provided abstracts during a qualitative analysis. However, when asked to identify specific quotes to illustrate themes, the GenAI tool generated fictional quotes, which, although they matched the themes identified, were not extracted directly from the source material. These challenges of hallucinations and a lack of transparency have previously been identified as a significant problem with using GenAI tools for qualitative research (Lee et al., 2024; Perkins & Roe, 2024a; Zhang et al., 2024), and although the use of analytic frameworks as a workaround has been suggested (Zhang et al., 2024), these issues are likely to remain for the near future. Therefore, one of the primary ethical concerns in using these tools for data analysis is the potential for misinterpretation or fabrication of research data.

**Quantitative Analysis**

Other ethical concerns may emerge when conducting quantitative data analyses. For example, GenAI tools have now developed to the point where they are able to perform forms of statistical analysis through the integration of Python libraries. This, combined with the speed of the tools and their ability to explain concepts that may be misinterpreted by less experienced researchers, may result in increases in instances of 'p-hacking', where statistical significance is present, but theoretical significance is lacking. Given that this is already a concern in academic research (Head et al., 2015), this could become more common as these tools are more widely used for research purposes.

The randomness and probabilistic decision-making nature of GenAI tools may also cause issues with the reproducibility of results (Perkins & Roe, 2024c) or alternate interpretations of data to suit a particular research goal, leading to additional confusion in the literature. As GenAI tools have the potential to repeat or enhance existing biases, even when they are not present in the data (Hacker et al., 2024), critically exploring the analysis provided by any GenAI tool is an essential step if these tools are integrated into the data analysis process.

**AI image generation**

Bendel (2023) evaluated three AI image generators (DALL-E 2, Stable Diffusion, and Midjourney) and found ethical issues such as copyright infringement (e.g. presence of watermarks and fonts in AI-generated images), disclosure of user privacy (e.g. prompt provided by user reveals user's interest or mindset), and lack of responsibility and liability (e.g. autonomous AI process unable to take moral and legal responsibility). The use of AI-generated scientific diagrams in peer-reviewed scientific articles has also resulted in retraction owing to inaccuracies and a lack of data integrity (Retraction Watch, 2024b).

Our exploration of a range of image generation platforms such as Midjourney, Dall-E-3, and ChatGPT revealed the presence of various ethical issues such as inaccurate representation of input texts in AI-generated images and videos, a possibility of copyright violation due to lack of information on the sources of images and videos, and lack of clarity about the protection and sharing of user data and input information.

Some platforms are recognising the growing backlash and concerns related to image generation and are making progress toward resolving some of these issues. For example, Public Domain 12m (Meyer et al., 2024) is a photographic database with synthetic captions designed for the ethical training of text-to-image models, and Adobe Firefly provides a detailed explanation of how the model is trained only on content on which it has permission (Adobe, 2024).

## Code Generation

Currently, there are many tools that create program code in a variety of programming languages based on prompt input. Single-purpose tools are available, either paid ones such as the GitHub Copilot or OpenAI Codex, or free ones such as Tabnine, Jedi, or CodeT5. Universal tools such as ChatGPT or Claude can also be used for this purpose. Use cases for these tools are often given as the automation of the "boring" or template parts of programming, for example, generating scripts for various data processing tasks, setting up the scaffolding for a programming task, or adapting code to another context. There are also so-called agent-based tools that integrate various AI tools throughout the software development and deployment life cycle, such as Replit or Cursor.

These tools enable programmers to concentrate more on addressing specific tasks by potentially reducing the need for detailed implementation considerations and minimising the time spent on code testing and debugging, particularly for errors arising from human factors, such as inattentiveness (Solohubov et al., 2023). ChatGPT-4 has demonstrated the ability to improve existing code quality metrics and generate tests with substantial coverage, though many tests fail when applied to the associated code (Poldrack et al., 2023).

Evaluations of popular AI coding assistants revealed varying levels of code correctness, with ChatGPT performing best at 65.2% compared to the benchmark HumanEval dataset (Yetiştiren et al., 2023). While these tools show promise, they still require human supervision to ensure accuracy and quality (Poldrack et al., 2023). AI tools can also be used to learn programming. One recent study showed that these tools significantly improved novice learners' code-authoring performance without negatively affecting their manual coding skills or retention (Kazemitabaar et al., 2023). Another study (Uplevel, 2024), however, found no significant time gains when comparing data from prior to the availability of Github Copilot with data from a similar time frame a year later. Uplevel also noted an increase of over 40% in the number of bugs that required fixing.

The ethical issues that we identify here start with potential licencing violations if the output reproduces a code that is under a non-free licence. The code produced, although often syntactically correct, may not actually solve the problem, as stated, with a number of iterations usually necessary to get the generated code to work as specified, thereby reducing the potential promised time savings. In particular, there needs to be test cases developed that comprise input and expected output before attempting to create code with such a system. The test suite must cover all possible edge cases. However, even when the code does finally work, gaping security holes may be found, and the generated code does not necessarily follow the Clean Code standards (Martin, 2008). Thus, additional work is necessary to clean up and harden the code prior to any use, which may not be carried out by all users of such tools, resulting in potential vulnerabilities in the published software.

## Academic Writing

### Grant proposal writing

Grant proposals represent collaborative institutional endeavours, typically involving multiple stakeholders including staff members, early career researchers, and professional writers working alongside the Principal Investigator. The integration of GenAI tools into this complex writing process requires careful consideration and potentially new guidelines from funding agencies (Meyer et al., 2023) to ensure research integrity. Although some funding bodies have begun to address how GenAI may be used in grant applications, their approaches vary significantly. For example, the Swedish Research Council is an agency that does not prohibit applicants from using GenAI tools in funding applications, stating, however, that applicants are responsible for the content of their applications and good research practice if these tools are used (Vetenskapsrådet, 2023).

While GenAI can be used to rewrite previous proposals, relying on GenAI for content creation may introduce hallucinations and downgrade the novelty of the research proposal. In addition, if the uploaded text is used to train future models, there is a risk of elements of the research proposal being incorporated into GenAI tools before the research has even started, potentially compromising the perceived originality of the work.

### Text generation

When considering text generation, several critical aspects must be evaluated to ensure the integrity and ethicality of the content produced. Both general GenAI tools such as ChatGPT, Copilot (institutional and personal), Gemini, and Grok in research, and specialised GenAI tools such as Kahubi, can generate different parts of a research paper. However, this raises significant epistemological and ethical concerns, particularly from the perspective of researchers working in potentially sensitive or controversial areas. One example is genocide studies, where AI-generated content may inadvertently trivialise or misrepresent facts, which can be profoundly harmful and disrespectful to survivors and descendants of genocide victims. Moreover, the potential misuse of AI-generated data for propaganda or revisionist purposes presents further ethical dilemmas, necessitating stringent content restrictions and careful supervision to prevent the dissemination of harmful and misleading information. The control of censorship within AI systems is fraught with challenges, including the secrecy surrounding guidelines and the potential for propaganda.

The 'black box' nature of AI decision-making exacerbates these concerns. As Resnik and Hossein argue, 'the opacity of AI systems is ethically problematic' (2023). Different LLMs may impose varying levels of content restrictions, with some models limiting responses on certain topics. These restrictions can lead to gaps in analysis and the spread of misinformation or disinformation, impacting academic freedom and preventing access to certain research areas (Waddington, 2024). AI-generated content often suffers from issues of accuracy and reliability, which are particularly problematic in a research context. The dissemination of misinformation or flawed conclusions can undermine the credibility of the research, leading to the spread of false narratives. For example, in the study of genocides, ensuring the factual accuracy of AI-generated content is of paramount importance, as historical inaccuracies not only misinform but also contribute to denialism or the minimisation of genocides. Given these risks, all AI-generated content must be meticulously checked for accuracy, a process that can be time-consuming and may counteract the perceived efficiency of using AI tools. While LLMs can provide useful contextual understanding, their ability to do so is not always consistent, necessitating the verification of both the facts presented and the contextual interpretation offered by these models. In cases where LLMs fail to grasp nuances, the resultant content may lack the depth and accuracy required for scholarly research.

The language and tone used when discussing sensitive topics must be handled with utmost sensitivity and respect, yet AI-generated content does not always adhere to these standards, risking the production of insensitive or offensive material. Although content restrictions can mitigate this risk, they also raise concerns, particularly when AI-generated content is perceived to 'lie' or misrepresent facts. The potential for bias in AI-generated text (as discussed in the literature review section) is a significant concern, particularly given that AI models can replicate or even amplify the biases present in the data on which they were trained. This can lead to the creation of biased or discriminatory content, which, if used in research, can perpetuate harmful stereotypes or misinform readers. Furthermore, if the training data lacks diversity, AI may generate content that inadequately represents certain groups, resulting in skewed or incomplete research findings. Conversely, there is also the risk of AI attempting to portray diversity where it has not existed historically. For example, in February 2024, Google issued an apology for 'missing the mark' after Gemini generated racially diverse 1943 German soldiers.

The question of authorship in AI-generated content is complex and raises significant concerns regarding academic integrity. If AI-generated research outputs lack clear attribution or contain 'hallucinated' sources, there is a risk of inadvertent plagiarism, as it is often unclear whether the AI is replicating the work of others verbatim. This uncertainty complicates the ethical use of AI in academic research and challenges the traditional notions of authorship and originality. The phenomenon of AI hallucinations where the system generates false or misleading content, poses significant challenges to research integrity, as they may introduce large datasets with false narratives, potentially leading to a rewriting of the 'truth', and the dissemination of inaccurate information.

**Text editing**

Various challenges, such as language barriers and some disabilities, can hinder individuals in text editing. However, although GenAI tools can help address these challenges, it is important to recognise its role as a supplementary tool not as an 'author'. In this paper, we have also used these tools to support text reduction and editing to achieve a more consistent voice throughout our manuscript, while ensuring that this process remained strictly under human oversight. However, it is important to be aware of potential biases that such tools may introduce, along with the risk of inaccuracies in summarising information or citing sources. Additionally, there is a risk of inaccuracies in the summarisation of information or citations of sources. There are also some caveats in the use of LLMs for editing text with an editor-based plug-in, as noted by Baron ( 2024). There is a tendency for the systems to redefine abbreviations repeatedly, use synonym substitutions that can cause "tortured phrases" (Else, 2021) to be used instead of standard terminology, many errors in the use of definite and indefinite articles in English, and occasional subject-verb agreement errors that occur. There also appears to be a tendency, Baron writes, for the rewriting engine to remove quotation marks from actual citation, lightly paraphrase the text and then remove the in-text citation, thus leading to plagiarism. This shows that while GenAI tools can streamline editing tasks, careful human input, verification, and critical judgment remain crucial

**Text proofreading**

Many universities in the UK are now allowing proofreading tools to be used in research but specify the caveat that the user be 'aware of the limitations of corrective software and generative AI, including translation tools. Over-reliance on digital tools can result in meaning being lost or distorted (Baron, 2024) and in a failure to convey appropriate understanding of the subject and the technical terms associated with the subject.' (University of Leeds, 2024). Researchers need to ensure that assistive tools do not alter the meaning of the text but instead focus on removing grammatical slips and ensuring logical coherence of the text.

**Translation**

GenAI can be used both to translate text for reading – thus enabling gathering literature and data content across language barriers – and also for the translation of the text written by a researcher to another language. Although both machine translation tools and LLM-based chatbots can be used to translate texts, they represent different types of technology and functionalities, thus partly raising different ethical concerns.

Machine translation tools, such as Google Translate, Microsoft Translator, and DeepL, are specifically created to translate text between different languages and are based on large databases of parallel corpora translated into multiple languages. While DeepL only supports 30 languages (DeepL, 2024), as of June 2024, Google Translate supports 243 languages (Google Blog, 2024), a development that was enabled by PALM 2, Google's AI model. AI has also been used to develop larger volumes of high-quality data for low-resource languages through transfer learning and data-mining in Meta's "No Language Left Behind" (NLLB) project which created a model that can handle 204 languages (Adelani, 2024; NLLB Team, 2024).

However, while able to produce translations, LLM-based chatbots, such as ChatGPT and BERT, are designed to generate any type of content. Having over 7.000 languages in the world (UNESCO, n.d.), the multilingualism of LLMs is also limited. In addition, LLMs can struggle with the accuracy of translation from less common languages; as the datasets are mainly in English, these models can encounter issues in representing cultural specificities (Tenzer et al., 2024) and specialised terms and writing styles (Gao et al., 2024). In addition, mixing of different standards of the same or similar languages is not uncommon, especially for low-resource languages, mixing standards of Bosnian, Croatian, and Serbian being one example found in our tests. GenAI datasets are trained on diverse Internet texts and, as such, include various biases on the structural level; those biases can affect the translation (Ghosh & Caliskan, 2023).

Although the use of AI detectors is not recommended owing to inherent problems with interpretation of the results, accuracy, and bias (Liang, 2023, Weber-Wulff et al., 2023, Perkins et al., 2024a, 2024b), it is worth noting that the reliability of the detectors significantly drops if machine translation is used on human written text, making it possible to wrongly accuse a researcher for using GenAI when using such tools for the translation of their own texts (Weber-Wulff et al., 2023).

## Peer Review and Ethical Publishing

Although GenAI may enable a faster peer review process (Mrowinski et al., 2017), compared to human reviewers, the current ability of these tools is limited (Suleiman et al., 2024). In addition, outsourcing this standard method for the evaluation of scientific quality to GenAI raises several ethical concerns over the reproduction and amplification of biases, lack of transparency, confidentiality of the data privacy, and sharing confidential information and unpublished research texts, as well as the reproducibility of the peer review (Hosseini & Horbach, 2023). Many journals and funding agencies also prohibit the use of GenAI in peer reviews or in the assessment of funding applications (Vetenskapsrådet, 2023). To address these concerns, oversight, accountability, and responsibility by peer reviewers and editors are required, including full disclosure.

The Committee on Publication Ethics (COPE) (2023) has published a position statement that GenAI tools cannot meet the requirements for authorship, as they cannot meet the criteria of being responsible for the work, nor can they sign copyright and licence agreements, nor can they assert potential conflicts of interest. However, they state that any use of GenAI tools should be disclosed.

Using GenAI in research poses ethical challenges concerning publication ethics enhancing the problems that already exist in the system of "publish or perish" that pressures researchers and institutions to focus on the quantity of outputs for career advancement. This environment incentivises unethical practices such as salami slicing (Šupak Smolčić, 2013), paper mills—businesses that produce academic papers for a fee (Committee on Publication Ethics & Scientific, Technical & Medical Publishers, 2022)— and predatory journals that publish research without rigorous peer review (Beall, n.d., Grudniewicz et al., 2019). The ability to produce papers of questionable scientific merit easily or automatically will potentially enhance such unethical practices (Kendall & Teixeira da Silva, 2024). At the same time, it should be noted that AI tools can also be used to combat such practices (Else, 2022).

# Discussion

The European Code of Conduct for Research Integrity states that good research practice is based on the fundamental principles of research integrity (Allea, All European Academies, 2023, p. 4):

- Reliability in ensuring the quality of research, reflected in the design, the methodology, the analysis and the use of resources.
- Honesty in developing, undertaking, reviewing, reporting and communicating research in a transparent, fair, full and unbiased way.
- Respect for colleagues, research participants, society, ecosystems, cultural heritage and the environment.
- Accountability for the research from idea to publication, for its management and organisation, for training, supervision and mentoring, and for its wider impacts.

Adherence to these principles and transparency in research practices are necessary when GenAI tools are used in research. Several ethical problems were encountered during the different research phases described in our study. The ethical challenges that were identified can be summarised as follows:

*Lack of transparency*: GenAI tools can be useful in many steps of the research process; however, as stated in *European Code of Conduct for Research Integrity*, concealing the use of AI tools in the content creation or drafting of publications is explicitly identifier as a violation of research integrity (Allea - All European Academies, 2023, p. 10). Researchers are required to report their methods, including the use of AI (Allea - All European Academies, 2023, p. 7), and disclose the use of AI in reviewing processes (Allea - All European Academies, 2023, p. 9).

*Copyright*: Researchers should be aware that uploading copyrighted material (e.g. research papers) to tools without authorisation can violate intellectual property laws.

*Privacy*: Ethical concerns regarding privacy in GenAI tools arise from the potential misuse of personal data that may be added to the system's dataset for further training, raising concerns over data protection. However, there are recent changes to handling the content which may lower the risk of misuse of personal data. For example, Microsoft Co-Pilot and ChatGPT have data privacy settings which do not use uploaded material to train future models, and the latest version of ChatGPT has an option to communicate via temporary chat where interactions are not stored or used for training, although a copy might be kept for up to 30 days.

Other ethical problems stem from the 'black box' nature of GenAI tools, which researchers cannot directly influence. Users of these tools cannot fully understand how the systems function, how underlying decisions are made, and what content might be censored or filtered, thus distorting the provided output. This lack of explainability makes it challenging to assess the validity of the generated output or address biases in the content. Therefore, researchers must consider this limitation when using GenAI in their studies.

*Inadvertent plagiarism and copyright breach*: Whenever GenAI is used to generate outputs, there is a risk that the content can inadvertently be plagiarised from an existing source. In addition, as GenAI tools are trained on datasets that include copyrighted material added without consent from the creators of that material, researchers who use such tools might inadvertently violate copyright laws by relying on GenAI outputs.

*Bias*: GenAI tools are trained on datasets that might reflect prejudices such as sexism, racism, and other stereotypes. Such biases might favour dominant perspectives and harm marginalised groups, perpetuating discrimination and inequalities. Additionally, bias in GenAI is amplified by a multiplier effect stemming from the systems being developed by programmers with Western perspectives. As the bias is included on the structural level, this can affect the way researchers read the literature, analyse the data, and interpret the data, and it might be reflected in the content generated by GenAI.

*Censorship*: Although several AI companies, such as OpenAI and Google, have tried to implement ethical post-training of their tools to reduce harmful or biased outputs, the inclusion of such training might limit the scope of research by restricting certain topics. Inability to generate output might thus interfere with academic freedom and open enquiry, posing barriers for researchers to explore certain topics.

*Fabrication (Hallucinations and False Data Sets)*: LLMs are trained to statistically predict content in a given context; thus, they sometimes produce fabricated, incorrect, or inaccurate information. Therefore, using tools based on LLMs means that researchers risk basing their findings on hallucinations (false information presented as factual), leading to false conclusions and undermining the integrity of scientific outputs.

The main ethical issues identified in different stages of research are summarised in Table 1.

*Table 1: Summary of ethical issues*

| Ethical Issue / Stage | Proposal writing | Formulating hypothesis and research questions and study design | Literature review and textual under-standing | Data collection | Data processing and analysis | Text creation, Translation, Transcription | Code generation | Publishing and peer review |
|---|---|---|---|---|---|---|---|---|
| **Lack of transparency** | X | X | X | X | X | X | X | X |
| **Copyright** | X | X | X | X | X | X | X | X |
| **Privacy** | X | X | X | X | X | X | | |
| **Inadvertent plagiarism and copyright breach** | X | X | X | X | X | X | X | X |
| **Bias** | X | X | X | X | X | X | X | X |
| **Censorship** | X | X | | X | | X | | X |
| **Fabrication (Hallucinations and false data sets)** | X | X | X | X | X | X | X | X |

While our research highlights certain ethical issues that might emerge during different stages of research, it is important to acknowledge that other ethical issues may also arise beyond those identified here. These issues might be context-specific and influenced by factors such as particular research settings or the GenAI tool used in the research process. Therefore, it is important that researchers remain prepared to address other ethical dilemmas that might arise when using GenAI in research.

Our findings indicate that various ethical issues can arise during the different phases of the research process. While most of these concerns remain consistently relevant, researchers must pay particular attention to specific ethical considerations during certain stages. This ensures that ethical standards are maintained, and that the integrity of the research process is upheld throughout.

## Recommendations for the ethical use of GenAI in research

To mitigate violations of these principles when using GenAI, we propose the following recommendations on the ethical use of GenAI in research (Table 2).

*Table 2: recommendations to deal with ethical issues*

| Ethical Issue | Allea principles violated | Recommendations |
| --- | --- | --- |
| **Lack of transparency** | **Honesty:** Failing to disclose the use of GenAI violates the principle of honesty by obscuring true origins of the work and preventing the assessment of its reliability. Researchers must disclose all methods and tools used. **Accountability:** Transparency is a key component of the principle of accountability. Researchers must take accountability for verifying the GenAI outputs. | If AI is involved in data analysis, content creation, or other substantive tasks, disclosure is necessary. If AI is used in non-substantive ways, e.g. only for minor text editing to improve the language of the text such as grammar and spelling, or for literature search, the information can be disclosed to be fully transparent. Researchers should transparently describe how GenAI was used in the methods and/or acknowledgements section. To document how GenAI has been used, researchers might want to note this in a research diary during the research process. It is not appropriate to attribute authorship to GenAI. |
| **Privacy** | **Respect:** Uploading personal data to GenAI tools violates the principle of respect for research participants. **Accountability:** Researchers are obliged to safeguard privacy, ensuring accountability for how data is collected, stored, and processed in GenAI. | To protect human subjects, avoid uploading confidential, sensitive, or personally identifiable data into GenAI to minimise the risk of data breaches. Develop protocols for managing data security. Ethical approval application and informed consent should include information that GenAI will be used for the data collection and/or analysis. Peer-reviewers should not upload manuscripts to GenAI tools for reasons of privacy and confidentiality. |
| **Copyright** | **Respect:** The use of copyrighted material without permission demonstrates a lack of respect for intellectual property rights. **Honesty:** Respecting copyright ensures honest and fair use of resources as well as compliance with legal and ethical standards. | Copyrighted content should not be uploaded to the GenAI tools without the copyright holder's permission. |
| **Inadvertent plagiarism and copyright breach** | **Honesty:** A lack of transparency in how GenAI generates the outputs conflicts with the principle of honesty and obscures the processes behind the research findings. GenAI tools may reproduce material from existing sources, which can lead to inadvertent plagiarism and inadvertent breach of copyright. | If GenAI content is directly quoted, reference it correctly. |
| **Bias** | **Reliability:** GenAI tools may provide incorrect answers due to their stochastic nature and inherent biases. This structural bias compromises the reliability of research outputs. **Respect:** Bias may also violate respect for research participants, as it can perpetuate stereotypes or marginalise certain groups. | To reduce potential bias, researchers should consider how they craft their input to GenAI and carefully evaluate and audit GenAI-generated outputs. While GenAI can be a valuable research tool, human oversight is essential to maintain good research practice, as researchers are ultimately responsible for all outputs of their work. |
| **Censorship** | **Respect:** GenAI tools might censor the outputs due to the output filters, which undermines respect for diverse perspectives and academic freedom. **Honesty:** Censorship contradicts the principle of honesty, preventing unbiased and full communication of research findings. | Human oversight is needed to ensure that GenAI tools do not exclude or suppress important research information. Deep discipline knowledge, as well as researchers' accountability are important to mitigate this ethical risk. |
| **Fabrication (Hallucinations and false data sets)** | **Honesty:** Fabrication directly challenges honesty in research by creating hallucinations and false or misleading data. **Reliability:** Fabrication impacts the reliability of the research process, and undermines the reliability of research outputs. | Validate all GenAI outputs by cross-referencing them with trusted sources. Human oversight and verification are needed at every stage. GenAI should not be used to create or manipulate research data and results. Generating images and blots using GenAI is therefore not recommended and should be approached with utmost caution and transparency. |

# Conclusion

Although AI-generated content offers potential benefits, particularly in terms of efficiency and accessibility, its application in research must be approached with caution. Ethical considerations, accuracy, bias, and the potential for misinformation are critical issues that must be carefully managed to ensure that GenAI serves as a tool for enhancing, rather than undermining, the integrity of research scholarships.

Our exploration of GenAI tools at different stages of the research process is subject to several limitations. We have not tested every available tool, and the tools are constantly evolving, meaning that our analysis may not capture the full scope of potential use and ethical challenges that may be encountered. Furthermore, although we sought to approach this study with objectivity, our biases influenced how we explored and evaluated the tools. That said, our team includes members from different cultural and disciplinary backgrounds, and with different perspectives of, and experiences with GenAI tools, which helps to mitigate these biases by offering different perspectives and interpretations.

As GenAI technologies and their use evolve, so too will perspectives on how researchers should use these tools and perform research. Further studies should involve conducting in-depth tests focused on specific disciplines or areas of research, as this report offers an overview rather than specialised insights. This iterative process will help refine the recommendations and understanding as the use of GenAI tools in research continues to develop.

### GenAI usage statement

This study used a range of GenAI tools to identify potential ethical issues as described throughout the manuscript. Claude 3.5 Sonnet was also used for revision and editorial purposes during the production of the manuscript. The authors reviewed, edited and take responsibility for all outputs of the tools used in this study.